\documentclass[letterpaper, onecolumn,10 pt]{IEEEtran}
\usepackage{microtype}
\usepackage{cite}
\usepackage{graphicx}
\usepackage{subfigure}
\usepackage[para]{footmisc}
\usepackage{booktabs} 
\usepackage{comment}
\usepackage{makecell}
\usepackage{boldline}
\makeatletter
\def\endthebibliography{%
  \def\@noitemerr{\@latex@warning{Empty `thebibliography' environment}}%
  \endlist
}
\makeatother

\usepackage{amsfonts,amsmath,amssymb}

\usepackage{amsthm}
\usepackage{mathtools}
\usepackage{multicol}
\usepackage{microtype}
\usepackage[table]{xcolor}
\newcommand{\PreserveBackslash}[1]{\let\temp=\\#1\let\\=\temp}
\newcolumntype{C}[1]{>{\PreserveBackslash\centering}p{#1}}
\usepackage{hyperref}
\usepackage{array}
\usepackage{tikz}
\newcolumntype{H}{>{\setbox0=\hbox\bgroup}c<{\egroup}@{}}

\newcommand{\Mod}[1]{\ \mathrm{mod}\ #1}
\newcommand{\sdr}{$\Delta$SDR}

\title{Wavesplit: End-to-End Speech Separation\\ by Speaker Clustering}

\author{%
  Neil Zeghidour, David Grangier \\
  Google Research, Brain Team\\
  \texttt{\{neilz, grangier\}@google.com} \\
  }

\begin{document}

\maketitle

\begin{abstract}
\normalsize{We introduce Wavesplit, an end-to-end source separation system. From a single mixture, the model infers a representation for each source and then estimates each source signal given the inferred representations. The model is trained to jointly perform both tasks from the raw waveform. Wavesplit infers a set of source representations via clustering, which addresses the fundamental permutation problem of separation. For speech separation, our sequence-wide speaker representations provide a more robust separation of long, challenging recordings compared to prior work. Wavesplit redefines the state-of-the-art on clean mixtures of 2 or 3 speakers (WSJ0-2/3mix), as well as in noisy and reverberated settings (WHAM/WHAMR). We also set a new benchmark on the recent LibriMix dataset. Finally, we show that Wavesplit is also applicable to other domains, by separating fetal and maternal heart rates from a single abdominal electrocardiogram.}
\end{abstract}

\section{Introduction}
\label{sec:intro}

Source separation is a fundamental problem in machine learning and signal processing, in particular in the ill-posed setting of separating multiple sources from a single mixture. An additional difficulty arises when the sources to separate belong to the {\it same} class of signals. For example, tasks such as separating overlapped speech, isolating appliance electric consumption from meter reading~\cite{murray2017electrical}, separating overlapped fingerprints~\cite{shehu2018sokoto}, identifying exoplanets in multi-planetary systems from light curves~\cite{shallue2018identifying} or retrieving individual compounds in chemical mixtures from spectroscopy \cite{toumi2014review} are particularly difficult as the sources are similar in nature, and as any permutation of them is a correct prediction. This leads to the fundamental {\it permutation problem} where predicted channels are well separated but inconsistent along time~\cite{weng15:energy}. This situation does not occur when separating different musical instruments from predefined categories~\cite{musdb18}, or separating speech from non-speech noise~\cite{loizou2013speech}. Thus, designing a model that maintains a consistent assignment between the ground-truth sources and the predicted channels is crucial for the tasks with similar sources.

This work precisely aims at separating sources of the same nature from a single mixture. In particular, speech separation aims at isolating individual speaker voices from a recording with overlapping speech~\cite{vincent18book}. This task is particularly important for public events, conversations and meeting recordings. Research on speech separation spans several decades~\cite{vincent18book} and it is the most active and competitive field of research in separation~\cite{hershey16:deep_clustering, luo19:tasnet, luo2019:dualpathrnn, nachmani2020voice}. We therefore introduce our model in the context of this application. Still, to show its generality, we also apply our model to the separation of fetal and maternal heart rate from a single abdominal electrode.

Our approach, Wavesplit, aims at separating novel sources at test time, called open speaker separation in speech, but leverages source identities during training. Specifically, our joint training procedure for speaker identification and speech separation differs from prior research~\cite{wang18:voicefilter}. The training objective encourages identifying instantaneous speaker representations such that (i) these representations can be grouped into individual speaker clusters and (ii) the cluster centroids provide a long-term speaker representation for the reconstruction of individual speaker signals. The extraction of an explicit, long-term representation per source is novel and is beneficial for both speech and non-speech separation. This representation limits inconsistent channel-source assignments (channel swap), a common type of error for permutation-invariant training (PIT), the dominant approach in neural source separation.

Our contributions are six-fold, (i) we leverage training speaker labels but do not need any information about the test speakers beside the mixture recording, (ii) we aggregate information about sources over the whole input mixture which limits channel-swap, (iii) we use clustering to infer source representations which naturally outputs sets, i.e. order-agnostic predictions, (iv) we report state-of-the-art results on the most common speech separation benchmarks, both for clean (WSJ0-{2/3}mix, Libri{2/3}mix clean) and noisy settings (WHAM and WHAMR, Libri{2/3}mix noisy), (v) we analyze the empirical advantages and drawbacks of our method, (vi) we show that our approach is generic and can be applied to non-speech tasks, by separating maternal and fetal heart rate from a single abdominal electrode.

\section{Related Work}
\label{sec:related}

Single-channel separation takes a single recording with overlapping source signals and predicts the isolated sources. This task is classical in speech processing~\cite{roweis01:one_mic,yilmaz04:blind,vincent18book} and has witnessed fast progress recently with supervised neural networks~\cite{wang18:overview}. These models have historically relied on learning time-frequency masks. They divide the input mixture in time-frequency bins (TFB) using a short-term Fourier Transform~\cite{williamson12:DSP}, and identify the source with the maximal energy for each TFB. Source spectrograms can then be produced by masking the TFBs of the other sources, and source signals are later estimated by phase reconstruction~\cite{griffinlim84,wang18:unfolded}. Soft masking variants assign each TFB to multiple speakers with different weights \cite{araki04:continuousmask}.
 
 {\bf Deep clustering approaches} devise a clustering model for masking: the model learns a latent representation for each TFB such that the distance between TFBs from the same source is lower than the distance between TFBs from different sources. The inference procedure clusters these representations to group TFBs by source \cite{hershey16:deep_clustering}. Wavesplit also relies on clustering to infer source representations but these representations are not tied to frequency bins and no masking is performed. Instead its representations are learned to (i) predict training speaker identity and (ii) provide conditioning variables to our separation convolutional network.

{\bf Permutation-Invariant Training (PIT)} avoids clustering and predicts multiple masks directly \cite{yu17:pit,kolbaek17:multitalker,xu18:pit}. The predictions are compared to the ground-truth masks by searching over the permutations of the source orderings. The minimum error over all permutations is used to train the model. PIT acknowledges that the order of predictions and labels in speech separation is irrelevant, i.e. separation is a set prediction problem. Among PIT systems, time domain approaches avoid phase reconstruction and its degradation. They predict audio directly with convolutional \cite{luo19:tasnet,zhang2020:furcanext} or recurrent networks \cite{luo2019:dualpathrnn}. In that case, PIT compares audio predictions to all permutations of the ground-truth signals. Wavesplit is also a time domain approach but it solves the permutation problem prior to signal estimation: at train time, the latent source representations are ordered to best match the labels prior to conditioning the separation network.

{\bf Discriminative speaker representations} can be extracted from short speech segments~\cite{wan18:verif,zeghidour16:siamese} to help separation.  
Wang et al. \cite{wang18:voicefilter} extract a representation of the targeted speaker from a clean enrollment sequence and then isolate that speaker in a mixture. This method however does not apply to open-set speaker separation, where enrollment data is not available for the test speakers. Nachmani et al. \cite{nachmani2020voice} use a separately trained speaker identification network to prevent channel swap. Conversely, Wavesplit jointly learns to identify and separate sources.

{\bf Separation of fetal and maternal heart rates} from abdominal electrocardiograms (ECGs) allows for affordable, non-invasive monitoring of fetal health during pregnancy and labour~\cite{bernardes2010persistent, karvounis2010non}, as an alternative to fetal scalp ECG. Source separation is an intermediate step for detecting fetal heart rate peaks~\cite{silva2013noninvasive, sutha2018fetal}. This work evaluates Wavesplit for separating maternal and fetal heart rate signals.

\section{Wavesplit}
\label{sec:wavesplit}

\begin{figure*} 
  \begin{center}
  \includegraphics[width=\textwidth]{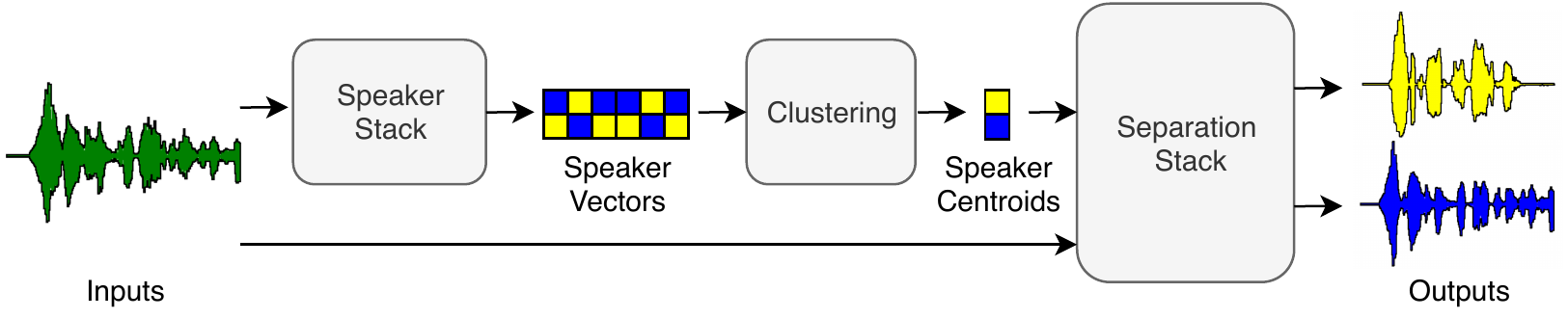}
  \end{center}
  \caption{Wavesplit for 2-speaker separation. The speaker stack extracts speaker vectors at each timestep. The vectors are clustered and aggregated into speaker centroids. The separation stack ingests the centroids and the input mixture to output two clean channels.}
  \label{fig:wavesplit}
 \end{figure*}

Wavesplit combines two convolutional subnetworks: the speaker stack and the separation stack (Figure~\ref{fig:wavesplit}). The speaker stack maps the mixture to a set of vectors representing the recorded speakers. The separation stack consumes both the mixture and the set of speaker representations from the speaker stack. It produces a multi-channel audio output with separated speech from each speaker.

The separation stack is classical and resembles previous architectures conditioned on
pre-trained speaker vectors~\cite{wang18:voicefilter}, or trained with PIT \cite{luo19:tasnet}. The speaker stack is novel and constitutes the heart of our contribution. This stack is trained jointly with the separation stack. At training time, speaker labels are used to learn a vector representation per speaker such that the inter-speaker distances are large, while the intra-speaker distances are small. At the same time, this representation is also learned to allow the separation stack to reconstruct the clean signals. At test time, the speaker stack relies on clustering to identify a centroid representation per speaker.

Our strategy contrasts with prior work. Unlike Wang et al. \cite{wang18:voicefilter}, we do not need an enrollment sequence for test speakers since the representation of all speakers is directly inferred from the mixture. With joint training, the speaker representation is not solely optimized for identification but also for the reconstruction of separated speech. In contrast with PIT~\cite{yu17:pit}, we condition decoding with a speaker representation valid for the whole sequence. This long-term representation yields excellent performance on long sequences, especially when the relative energy between speakers is varying, see Section~\ref{sec:clean_setting}. Our model is less prone to channel swap since clustering assigns a persistent source representation to each channel. Still in contrast with PIT, we resolve the permutation ambiguity during training at the level of the speaker representation, i.e. the separation stack is conditioned with speaker vectors ordered consistently with the labels. This does not force the separation stack to choose a latent ordering and allows training this stack with different permutations of the same labels.

\subsection{Problem Setting \& Notations}
\label{sec:problem_setting}

We consider a mixture of $N$ sources. Each single-channel source waveform $i \in [1,N]$ is represented 
by a continuous vector $y^i \in \mathcal{X}^{1,T}$, with $T$ the length of the sequence. Given a mixture \mbox{$x = \sum_{i=1}^N y^i$}, the source separation task is to reconstruct each $y^i$.

A separation model $f$ predicts an estimate for each channel. Its 
quality is assessed by comparing its predictions to the reference channels 
$\{y^i\}^N_{i=1}$ up to a permutation since the channel order is arbitrary,
\begin{equation}\label{eq:permutation_quality}
Q(\hat{y}, y) = \max_{\sigma \in S_N} \frac{1}{N} \sum_{i=1}^N q(\hat{y}^{\sigma(i)}, y^{i})
\quad
\textrm{where}
\quad
\forall i,
\hat{y}^i = f^i(x).
\end{equation}
$q(\cdot, \cdot)$ denotes a single-channel reconstruction quality metric and
$S_N$ denotes the space of permutations over $[1, N]$.
The speaker separation literature typically relies on Signal-to-Distortion Ratio (SDR) to
assess reconstruction quality. SDR is the opposite of the log squared error normalized by the energy of the reference signal,
\mbox{
${\rm SDR}(\hat{y}, y) = - 10 \log_{10}\left(\|y - \hat{y}\|^2 \right)  + 10 \log_{10} \left(\|y\|^2\right)$.
}
Scale-invariant SDR (SI-SDR) considers prediction scale irrelevant and searches over gains~\cite{leroux19:sdr}. Variants searching over richer signal transforms have also been proposed~\cite{vincent06:bss}.

\subsection{Model Architecture}
\label{sec:modelarch}

Wavesplit is a residual convolutional network with two sub-networks or {\it stacks}. The first stack transforms 
the input mixture into a representation of each speaker, while the second stack transforms the input mixture into multiple isolated recordings conditioned on the speaker representation.

The {\bf speaker stack} produces speaker representations at each time step and then performs an aggregation over the whole sequence. Precisely, the speaker stack first maps the input $x = {x}_{t=1}^{T}$ into $N$ same-length sequences of latent vectors of dimension $d$, i.e. $h(x) = \{h^i\}^N_{i=1}$ where $\forall i, h^i \in \mathbb{R}^{T \times d}$.
$N$ represents the maximum number of simultaneous speakers targeted by the system, while 
$d$ is a hyper-parameter selected by cross-validation. Intuitively, $h$ produces a latent representation of each speaker at every time step. It is important to note that $h$ is not required to order speakers consistently across a sequence. E.g. a  given speaker Bob could be represented by the first vector $h^1_t$ at time $t$ and by the second vector $h^2_{t'}$ at a different time $t'$.
At the end of the sequence, the aggregation step
groups all vectors by speaker and outputs $N$ summary vectors for the whole sequence.
K-means clustering performs this aggregation at inference~\cite{lindebg80:kmeans} and returns the centroids 
of the $N$ identified clusters,
\[
c = \{c_i\}^N_{i=1} = {\rm kmeans}(\{h^i_t\}_{i,t}; N).
\]
In the following, we refer to the local vectors $h^i_t$ as the {\it speaker vectors}, and to the vectors $c_i$ as the {\it speaker centroids}. During training, clustering is not used. Speaker centroids are derived by grouping speaker vectors by speaker identity, relying on the speaker training objective described in Section \ref{sec:objective}.

The {\bf separation stack} maps the mixture $x$ and the speaker centroids $c$ into an $N$-channel signal $\hat{y}$,
\[
\hat{y} = f(x, c) = f(x, {\rm kmeans}(h(x); N)).
\]

Inspired by Luo and Mesgarani \cite{luo19:tasnet}, we rely on a {\bf residual convolutional architecture} for both stacks. Each residual block in
the speaker stack composes a dilated convolution ${\rm dconv}$~\cite{yu16:dilated}, a non-linearity $\rm nl$ and layer normalization $\rm lnorm$~\cite{ba2016:layernorm},
\[
x_{l+1} = x_{l} + {\rm lnorm(}{\rm nl}({\rm dconv}(x_l))).
\]
We use parametric rectified linear units~\cite{he15prelu} for $nl$ after experimenting with multiple alternatives.
The last layer of the speaker stack applies Euclidean normalization to the speaker vectors.

The residual blocks of the {\bf separation stack} are conditioned by the speaker centroids relying on FiLM, Feature-wise Linear Modulation~\cite{perez18film},
\[
x_{l+1} = x_{l} + {\rm lnorm(}{\rm nl}(a * {\rm dconv}(x_{l}) + b))
\]
where $a = {\rm lin}(c)$ and $b= {\rm lin'}(c)$ are different linear projections of $c$, the concatenation of the speaker centroids. 
Section \ref{sec:ablation} shows the advantage of FiLM conditioning over classical bias only conditioning, $a=1$~\cite{oord:wavenet}.
We learn distinct parameters for each layer for all parametric functions, 
i.e. ${\rm dconv}$, $\rm nl$, $\rm lnorm$, $\rm lin$ and $\rm lin'$. 

\subsection{Model Training Objective}
\label{sec:objective}

Model training addresses two objectives: 
(i) it learns speaker vectors which can be clustered by speaker identity into well separated clusters;
(ii) it optimizes the reconstruction of the separated signals from aggregated speaker vectors.

Wavesplit assumes the training data is annotated with speaker identities from a finite set of $M$ training speakers but does not require any speaker annotation at test time. Speaker identities are labeled in most separation datasets, including Wall Street Journal variants~\cite{hershey16:deep_clustering,wichern19:WHAM,maciejewski19:whamr}, meeting recordings~\cite{mccowan05:ami} or cocktail party recordings~\cite{barker18:chime5}, but mostly unused in the source separation literature. Wavesplit exploits this information to build an internal model of each source and improve long-term separation.

The {\bf speaker vector objective} encourages the speaker stack outputs to have small intra-speaker and large inter-speaker distances. 
From an input $x$ with target signals $\{y^i\}^N_{i=1}$ and corresponding speakers $\{s_i\}^N_{i=1} \in [1, M]^N$, 
the speaker loss favors correct speaker identification at each time step $t$, i.e.
\[
{\cal L}_{\rm speaker}(x, \{s_i\}^N_{i=1})
= \sum_{t=1}^{T} \min_{\sigma \in S_N} \sum_{i=1}^N {\ell}_{\rm speaker}(h^{\sigma(i)}_t, s_i)
\]
where ${\ell}_{\rm speaker}$ defines a loss function between a vector of $\mathbb{R}^d$
and a speaker identity of $[1, M]$. The minimum over permutations expresses that each identity should be identified at each time step, in any arbitrary order. 
The best permutation (${\rm argmin}$) at each time-step is used to re-order the speaker vectors in an order consistent with the training labels. This allows averaging the speaker vectors originating from the same speaker at training time. This makes optimization simpler compared to work relying on k-means during training~\cite{hershey17:deepunfold}. This permutation per time-step differs from PIT~\cite{xu18:pit}: we do not require the model to pick a single ordering over the output channels, as eventual channel swaps at the level of speaker vectors will be corrected by k-means. Moreover, the separation stack is trained from different permutations of the same labels.

Three alternative are explored for ${\ell}_{\rm speaker}$. All three maintain
an embedding table over training speakers $E \in \mathbb{R}^{M \times d}$. First, 
${\ell}^{\rm dist}_{\rm speaker}$ is a distance objective. This loss
favors small distances between a speaker vector and the corresponding embedding
while enforcing the distance between different speaker vectors at the same time-step to 
be larger than a margin of 1,
\begin{equation}
{\ell}^{\rm dist}_{\rm speaker}(h^{j}_t, s_i) = 
  \| h^{j}_t - E_{s_i} \|^2  
+ \sum_{k \neq j} \max(0, 1 - \| h^{j}_t - h^{k}_t \|^2)
\label{eq:distance_loss}
\end{equation}
Second, ${\ell}^{\rm local}_{\rm speaker}$ is a local classifier objective
which discriminates among the speakers present in the
sequence. It relies on the log softmax over the distances
between speaker vectors and embeddings,
\begin{equation}
{\ell}^{\rm local}_{\rm speaker}(h^{j}_t, s_i) = 
d(h^{j}_t, E_{s_i}) + \log \sum_{k=1}^{N} \exp\left( -d(h^{j}_t, E_{s_k}) \right)
\label{eq:local_classifier}
\end{equation}
where $d(h^{j}_t, E_{s_i}) = \alpha \| h^{j}_t - E_{s_i} \|^2 + \beta$
is the squared Euclidean distance rescaled with learned scalar parameters $\alpha > 0, \beta$.
Finally, the global classifier objective ${\ell}^{\rm global}_{\rm speaker}$ is similar,
except that the partition function is computed over all speakers in the training set,
i.e. 
\begin{equation}
{\ell}^{\rm global}_{\rm speaker}(h^{j}_t, s_i) = 
d(h^{j}_t, E_{s_i}) + \log \sum_{k=1}^{M} \exp \left( -d(h^{j}_t, E_{k}) \right).
\label{eq:global_classifier}
\end{equation}
We use ${\cal L}_{\rm speaker}$ to update the speaker stack, as well as the speaker embedding table.
The {\bf reconstruction objective} aims at optimizing the separation quality, as 
defined in Eq.~(\ref{eq:permutation_quality}). 
\begin{equation}
{\cal L}_{\rm reconstr}(f(x, c), y) = \frac{1}{N} \sum_{i=1}^N {\ell_{\rm reconstr}}(f(x, c)^i, y^{i}).
\label{eq:reconstr_loss}
\end{equation}
Contrasting with PIT~\cite{yu17:pit}, this expression 
does not require searching over the space of permutations since the centroids $c = \{c_i\}_{i=1}^N$ are 
consistent with the order of the labels $\{y^{i}\}_{i=1}^N$ as explained above.
For $\ell_{\rm reconstr}$, we use negative SDR with a clipping 
$\tau$ to limit the influence of the best training predictions,
\mbox{$
\ell_{\rm reconstr}(\hat{y}, y) = - \min(\tau, {\rm SDR}(\hat{y}, y)).
$}
Inspired by~\cite{nachmani2020voice}, we compute ${\cal L}_{\rm reconstr}$ at each layer of the separation
stack, and use the average over all layers as our reconstruction loss.

We consider different forms of {\bf regularization} to improve generalization to new speakers. At training time, we add Gaussian noise to speaker centroids, we replace full centroids with zeros (speaker dropout) and we replace some centroids with a linear combination with other centroids from the same training batch (speaker mixup). Speaker dropout and mixup are inspired by dropout~\cite{srivastava14:dropout} and mixup~\cite{zhang17:mixup}. Note that regardless of the number of speakers in a sequence, speaker dropout removes at most one centroid, such that the separation task is not ambiguous ($N-1$ speaker-channel assignments are sufficient to reconstruct the $N$th). Finally, we favor well separated embeddings for the training speakers with entropy regularization~\cite{sablayrolles19spreading},
\mbox{$
\ell_{\rm reg} = -\sum_{i} \min_{j \neq i} \log \|E_{i} - E_{j}\|.
$}
Section \ref{sec:ablation} analyses the benefits of regularization. 

\subsection{Training Algorithm}

Model training optimizes the weighted sum of ${\cal L}_{\rm speaker}$ and ${\cal L}_{\rm reconstr}$ with Adam~\cite{kingma14:adam}. We train on mini batches of fixed-size windows. Wavesplit performs well for a wide-range of window sizes starting at 750ms, unlike most PIT approaches~\cite{luo19:tasnet, luo2019:dualpathrnn, nachmani2020voice} that require longer segments ($\sim$ 4s).
The training set is shuffled at each epoch and a window starting point is uniformly sampled each time a sequence is visited. This sampling gives the same importance to each sequence regardless of its length. This strategy is consistent with the averaging of per-sequence SDR used for evaluation. We replicate each training sequence for all permutations of the target signals to avoid over-fitting a specific ordering. Replication, windowing and shuffling are not applied to validation or test data.

\subsection{Data Augmentation with Dynamic Mixing}

Separation benchmarks
like WSJ0-2mix~\cite{hershey16:deep_clustering} create a standard split between train, valid and test sequences and then generate a finite set of input mixtures by summing specific clean signals with specific weights (gains). As an orthogonal contribution to our model, we consider creating training mixtures dynamically. Our training augmentation creates new examples indefinitely by sampling random windows of training recordings to be summed after applying random gains. A similar method has been used in music source separation \cite{uhlich17:musicaug, defossez19:musicaug}. This simple method brings systematic improvements, which advocate for the use of this training scheme when generating mixtures on the fly is not costly. We also experiment without augmentation to isolate the impact of Wavesplit alone.

\begin{table*}[t]
        \small
        \centering
        \caption{SI-SDR and SDR improvements (dB) on WSJ0-2mix and WSJ0-3mix.}
        \vspace{0.2cm}
        \label{table:wsj0_2_3mix_improv}
        \begin{tabular}{l|c|c|c|c}
                \hline\hline
                \multicolumn{1}{c|}{Model} & \multicolumn{2}{c|}{2 speakers} & \multicolumn{2}{c}{3 speakers} \\
                \multicolumn{1}{c|}{} & $\Delta$SI-SDR & $\Delta$SDR  & $\Delta$SI-SDR & $\Delta$SDR  \\
                \hline\hline
                Deep Clustering \cite{isik16:deep_clustering}            & 10.8 & -- & 7.1 & -- \\
                uPIT-blstm-st \cite{kolbaek17:multitalker}               &  --  & 10.0 & -- & 7.7 \\
                Deep Attractor Net. \cite{chen17:deep_attractor}         & 10.5 & -- & 8.6 & 8.9  \\
                Anchored Deep Attr. \cite{luo18:deep_attractor}          & 10.4 & 10.8 & 9.1 & 9.4 \\
                Grid LSTM PIT \cite{xu18:pit}                            &  --  & 10.2 & -- & --  \\
                ConvLSTM-GAT \cite{li18:cbldnn}                          &  --  & 11.0 & -- & --  \\
                Chimera++ \cite{wang18:deep_clustering_objectives}       & 11.5 & 12.0 & -- & --  \\
                WA-MISI-5 \cite{wang18:unfolded}                         & 12.6 & 13.1 & -- & --  \\
                blstm-TasNet \cite{luo18:tasnet}                         & 13.2 & 13.6 & -- & --  \\
                Conv-TasNet \cite{luo19:tasnet}                          & 15.3 & 15.6 & 12.7 & 13.1 \\
                Conv-TasNet+MBT \cite{lam19:mixupbreakdown}              & 15.5 & 15.9 & -- & --  \\
                DeepCASA \cite{liu19:deepcasa}                           & 17.7 & 18.0 & -- & -- \\
                FurcaNeXt \cite{zhang2020:furcanext}                     &  --  & 18.4 & -- & -- \\
                DualPathRNN \cite{luo2019:dualpathrnn}                   & 18.8 & 19.0 & -- & -- \\
                Gated DualPathRNN \cite{nachmani2020voice}               & 20.1 & -- & 16.9 & --  \\           
                \hline
                Wavesplit                                                &\bf{21.0}&\bf{21.2} & \bf{17.3} & \bf{17.6} \\
                Wavesplit + Dynamic mixing                               &\bf{22.2}&\bf{22.3} & \bf{17.8} & \bf{18.1} \\
                \hline                
        \end{tabular}
\end{table*}
\section{Experiments \& Results}
\label{sec:experiments}

Most experiments are performed on the speaker separation dataset~\cite{wsj0_2mix_dataset}
built from the LDC WSJ-0 dataset \cite{garofolo93:wsj0} as introduced in \cite{hershey16:deep_clustering}. We rely on the 8kHz version of the data, with 2 or 3 concurrent speakers. This setting is a de-facto benchmark for open-speaker source 
separation and we compare our results to alternative methods. Appendix \ref{appendix:wsj0mix_stats} reports
the dataset statistics. Additionally, we perform experiments in noisy settings. We rely
on WHAM! with urban noise~\cite{wichern19:WHAM} and WHAMR! with noise and reverberation~\cite{maciejewski19:whamr}. 
These datasets are derived from WSJ0-2mix and have identical statistics.
We further evaluate variants of Wavesplit with different loss functions and architectural alternatives. 
We conduct an error analysis examining a small fraction of sequences with a strong 
negative impact on overall performance. We also perform experiments on the recently released LibriMix dataset~\cite{librimix}. We also rely on the 8 kHz version of the data. Like for WSJ-0, we evaluate our model in clean and noisy settings with 2 or 3 concurrent speakers. The statistics of this larger dataset are also in Appendix. Finally, we show results on a fetus/mother heart rate separation task.

Our evaluation uses signal-to-distortion ratio (SDR) and scale-invariant 
SDR (SI-SDR)~\cite{vincent06:bss,leroux19:sdr}, see Section~\ref{sec:problem_setting}. 
SDR is measured using the standard MIR-eval library~\cite{mir_eval}.
Like prior work, we report results as \textit{improvements}, i.e. the metric obtained using the system 
output minus the metric obtained by using the input mixture as the prediction.  We provide recordings processed by our system on a public webpage \footnote{\url{https://soundcloud.com/wavesplitdemo/sets}} as well as in supplementary material.

\subsection{Hyperparameter Selection}

Preliminary experiments on WSJ0-2mix \cite{hershey16:deep_clustering} drove our architecture choices for subsequent experiments. 
Both stacks have a latent dimension of $512$ and the dilated convolutions have a kernel size of 3 without striding, therefore all activations preserve the temporal resolution of the input signal and no upsampling is necessary at the output layers. The dilation factor varies with depth. The speaker stack is $14$-layer deep and dilation grows exponentially from $2^0$ to $2^{13}$. The separation stack has $40$ layers with the dilation pattern from Oord et al. \cite{oord:wavenet}, i.e. $\delta_l = 2^{l \Mod{10}}$. Every 10 layers, the dilation is reset to 1 allowing multiple fine-to-coarse-to-fine interactions across the time axis.

For training, we validated a learning rate of $1\mathrm{e-}3$ in $[1\mathrm{e-}3, 2\mathrm{e-}3, 3\mathrm{e-}3]$ and a speaker loss weight of $2$ in $[1, 2, 5]$. For regularization, we validated a distance regularization weight at $0.3$ in $[0, 0.2, 0.3, 0.5]$ and a Gaussian noise with standard deviation at $0.2$ in $[0, 0.1, 0.2, 0.3]$. We use a speaker dropout rate of $0.4$ (picked in $[0, 0.2, 0.4, 0.6]$) and a speaker mixup rate of $0.5$ (picked in $[0, 0.5, 1]$). The clipping threshold on the negative SDR loss was validated at $30$ for clean data and $27$ for noisy data within $[22, 27, 30]$.
For ${\cal L}_{\rm speaker}$, we found the global classifier to be the most effective, see Section~\ref{sec:ablation}.

\subsection{Clean Settings}
\label{sec:clean_setting}

WSJ0-2mix/3-mix is the de facto benchmark for separation.
Table~\ref{table:wsj0_2_3mix_improv} reports the results for 2 and 3 simultaneous speakers. In both cases, Wavesplit outperforms alternatives
and dynamic mixing further increases this advantage. For instance, on WSJ0-2mix we report $21.0$ $\Delta$SI-SDR compared to $20.1$ for the recent gated dual path RNN~\cite{nachmani2020voice}. This number improves to $22.2$ with dynamic augmentation.

On WSJ0-2mix, Table \ref{table:error_analysis} analyses the error distribution. The test set has more sentences with poor \sdr{} ($<10$) compared to validation, $5.6\%$ versus $0.9\%$. Unlike test data, validation data contains the same speakers as the training set and we observe that test examples with low \sdr{} are sequences where both speakers are close to the same training speaker identity, according to the learned embeddings (confusing speaker). Our oracle permutes the predicted samples across channels, and reports the best permutation, showing that most of the errors are channel assignment errors.

WSJ0-2mix recordings have a single dominant speaker, i.e. the same speaker stays
the loudest throughout the whole recording. PIT might implicitly rely on this bias
to address the channel assignment ambiguity and we evaluate robustness to change in dominant speaker on long sequences. We concatenate test sequences with the same pair of speakers, for length up to 10 times the original length. Unlike the training data, the loudest speaker changes between each concatenated sequence. For PIT models, we retrain Conv-TasNet~\cite{convtasnet_github} and take a pre-trained Dual-Path RNN model~\cite{dprnn_github}. Table~\ref{table:wsj0_long} shows the advantage of Wavesplit, which explicitly models sources in a time-independent fashion rather than relying on implicit rules that exploit biases in the training data. This is remarkable since Wavesplit is trained on 1s long windows compared to longer 4s windows for both PIT models.

\begin{table}[t]
    \small

    \parbox{.50\linewidth}{
        \centering
        \caption{Error analysis on WSJ0-2mix.}
        \vspace{0.2cm}
        \label{table:error_analysis}
        \begin{tabular}{l V{3} r | r V{3} r | r}
                \hline\hline
                                                    & \multicolumn{2}{c V{3}}{valid} & \multicolumn{2}{c}{test}\\
                {\footnotesize $\Delta$SDR Split}   & $< 10$ & $\geq 10$ & $< 10$ & $\geq 10$ \\\hline\hline
                {\footnotesize Examples \%}         &   0.9  &  99.1 &  5.6 & 94.4\\
                {\footnotesize Confusing spkr \%}   &  60.7  &   5.5 & 77.2 & 12.4\\
                {\footnotesize Mean $\Delta$SDR}    &   4.3  &  22.4 &  3.9 & 22.2\\ %
                {\footnotesize Oracle $\Delta$SDR}  &  17.4  &  22.9 & 18.8 & 22.9 \\\hline %
                \hline
        \end{tabular}
    }
    \hfill
    \parbox{.40\linewidth}{
        \centering
        \caption{$\Delta$SDR (dB) on long sequences.}
        \vspace{0.2cm}
        \label{table:wsj0_long}
        \begin{tabular}{l|c|c|c}
                \hline\hline
                \multicolumn{1}{c|}{Model} & \multicolumn{3}{c}{Sequence Length}\\
                                & $\times 1$ &  $\times 4$ &  $\times 10$ \\
                \hline\hline
                Conv-TasNet    & 15.6 & 13.6 & 14.0 \\
                DualPathRNN    & 19.1 & 17.3 & 16.9 \\\hline
                Wavesplit      & {\bf 21.2} & {\bf 20.2} & {\bf 20.0} \\
                \hline
        \end{tabular}   
    }
\end{table}

\begin{table}[t]
    \small
    \centering
    \caption{SI-SDR and SDR improvements (dB) on WHAM! and WHAMR!.}
    \vspace{0.2cm}
    \label{table:wsj0_wham_whamr}
    \begin{tabular}{l|c|c|c|c}
            \hline\hline
            \multicolumn{1}{c|}{Model} & \multicolumn{2}{c|}{WHAM!} & \multicolumn{2}{c}{WHAMR!} \\
            \multicolumn{1}{c|}{} & $\Delta$SI-SDR & $\Delta$SDR  & $\Delta$SI-SDR & $\Delta$SDR  \\
            \hline\hline
            Conv-TasNet \cite{pariente19:filterbankDF, maciejewski19:whamr}              & 12.7 & -- & 8.3 & --\\
            Learnable fbank \cite{pariente19:filterbankDF}          & 12.9 & -- & -- & --\\            
            BLSTM-TasNet \cite{maciejewski19:whamr}     & 12.0 & -- & 9.2 & -- \\\hline
            Wavesplit                                                    &\bf{15.4}&\bf{15.8}& \bf{12.0} & \bf{11.1} \\
            Wavesplit + Dynamic mixing                                   &\bf{16.0}&\bf{16.5}& \bf{13.2} & \bf{12.2} \\
            \hline
    \end{tabular}
\end{table}

\begin{table*}[t]
    \small
    \centering
    \caption{SI-SDR and SDR improvements (dB) on LibriMix.}
    \vspace{0.2cm}
    \label{table:librimix}
    \begin{tabular}{l|c|c|c|c|c|c|c|c}
            \hline\hline
            \multicolumn{1}{c|}{Model} & \multicolumn{4}{c|}{Libri2mix} & \multicolumn{4}{c}{Libri3mix} \\\hline
            \multicolumn{1}{c|}{Condition} & \multicolumn{2}{c|}{clean} & \multicolumn{2}{c}{noisy} & \multicolumn{2}{c|}{clean} & \multicolumn{2}{c}{noisy} \\
            \multicolumn{1}{c|}{} & $\Delta$SI-SDR & $\Delta$SDR  & $\Delta$SI-SDR & $\Delta$SDR & $\Delta$SI-SDR & $\Delta$SDR & $\Delta$SI-SDR & $\Delta$SDR \\
            \hline\hline
            Conv-TasNet \cite{librimix}    & 14.7 & -- & 12.0 & -- & 12.1 & -- & 10.4 & -- \\
            IRM (oracle) \cite{librimix}            & 12.9 & -- & 12.0 & -- & 13.1 & -- & 12.6 & -- \\
            IBM (oracle) \cite{librimix}            & 13.7 & -- & 12.6 & -- & 13.9 & -- & 13.3 & -- \\\hline            
            Wavesplit                       & 19.5 & 20.0 & 15.1 & 15.8 & 15.8 & 16.3 & 13.1 & 13.8 \\
            Wavesplit + Dynamic mixing      & \bf{20.5} & 20.9 & \bf{15.2} & 15.9 & \bf{17.5} & 18.0 & \bf{13.4} & 14.1 \\
            \hline
    \end{tabular}
\end{table*}

\subsection{Noisy and Reverberated Settings}

WSJ0-2mix was recorded in clean conditions and noisy variants have been introduced to devise more challenging use cases. WHAM! \cite{wichern19:WHAM} adds noise recorded in public areas to the mixtures. As the model should only predict clean signals, it cannot exploit the fact that predicted channels should sum to the input signal. WHAMR! \cite{maciejewski19:whamr} adds the same noise, but also reverberates the clean signals. The task is even harder as the model should predict clean signals without reverberation, i.e. jointly addressing denoising, dereverberation and source separation. Table \ref{table:wsj0_wham_whamr} shows that our model outperforms previous work by a substantial margin. We also adapted dynamic mixing for these datasets. For WHAM!, we also sampled a gain for the noise, and combined it to reweighted clean signals to generate noisy mixtures on the fly. We similarly remixed WHAMR!, except that we reweighted reverberated signals with noise. On both datasets, this leads to an even larger improvement over previous work: e.g. our accuracy on WHAMR! is comparable to results on clean inputs (WSJ0-2mix) prior to~\cite{wang18:deep_clustering_objectives}.

\subsection{Large scale experiments on LibriMix}

We train Wavesplit on the newly released dataset LibriMix \cite{librimix}, which contains artificial mixtures of utterances from Librispeech~\cite{panayotov2015librispeech}, in four conditions (2/3 speakers, clean or noisy). We train on the \textit{train-360} subset of LibriMix to allow a fair comparison with the baselines from \cite{librimix}, which are trained in the same conditions. Like for WSJ-0, we rely on the 8 kHz version of the data. LibriMix is a good testbed for Wavesplit, as its training set contains significantly more speakers than WSJ0-2/3mix (921 in \textit{train-360} against 101 in the training set of WSJ0-{2/3}mix) which improves the robustness of the speaker stack. Table \ref{table:librimix} reports the results on all four conditions. The baselines are a Conv-TasNet model, as well as frequency masking oracles, Ideal Ratio Mask (IRM) and Ideal Binary Mask (IBM). Wavesplit significantly outperforms Conv-TasNet in all conditions, and even consistently outperforms oracle ideal masks. Moreover, the results are in the same range as the equivalent condition in WSJ0-{2/3}mix and WHAM!, which again confirms the robustness of the method across datasets.

\subsection{Ablation Study}
\label{sec:ablation}

Table~\ref{table:wsj0_ablation} compares the base result (no dynamic mixing) obtained with the global classifier loss, Eq.~(\ref{eq:global_classifier}), with the distance loss, Eq.~(\ref{eq:distance_loss}). Although this type of loss is common in distance learning for clustering~\cite{wang18:deep_clustering_objectives}, the global classifier reports better results. We also experimented with the local classifier loss, Eq.~(\ref{eq:local_classifier}), which yielded slower training and worse genereralization.
Table~\ref{table:wsj0_ablation} also reports the advantage of multiplicative FiLM conditioning compared to standard additive conditioning~\cite{oord:wavenet}. Not only reported SDRs are better but FiLM allows using a higher learning rate and yields faster training. Table~\ref{table:wsj0_ablation} also shows the benefit of regularizing the speaker representation.

\begin{table}[t]
    \small
    \parbox{0.45\linewidth}{
        \centering
        \caption{Ablation on WSJ0-2mix.}
        \vspace{0.2cm}
        \label{table:wsj0_ablation}
        \begin{tabular}{l|c}
                \hline\hline
                \multicolumn{1}{c|}{Model} & $\Delta$SDR (dB)  \\
                \hline\hline
                Base model           & 21.2 \\\hline
                w/ distance loss     & 19.3 \\
                w/o FiLM             & 20.6 \\
                w/o distance reg.    & 20.6 \\
                w/o speaker dropout  & 20.6 \\
                w/o speaker mixup    & 19.6 \\
                \hline
                \end{tabular}
                }
    \hfill
    \parbox{0.45\linewidth}{
        \centering
        \caption{SI-SDR and SDR improvements (dB) on FECGSYNDB (fetal/maternal ECG).}
        \vspace{0.2cm}
        \label{table:ecg}
        \begin{tabular}{l|c|c}
                \hline\hline
                \multicolumn{1}{c|}{Model} & $\Delta$SI- & $\Delta$SDR \\
                \multicolumn{1}{c|}{} & SDR &  \\
                \hline\hline
                Conv-TasNet & 11.4 & 11.9 \\
                DualPathRNN & 11.4 & 11.4 \\\hline
                Wavesplit   &\textbf{12.3}&\textbf{14.4}\\
                \hline
                \end{tabular}
                }
\end{table}

\subsection{Separation of Maternal and Fetal Electrocardiograms}

Wavesplit separation method can be applied beyond speech. Electrocardiogram (ECG) reports voltage time series of the electrical activity of the heart from electrodes placed on the skin. During pregnancy, ECG informs about the function of the fetal heart but maternal and fetal ECG are mixed. We aim at separating these signals from a single noisy electrode recording. We use the FECGSYNDB data \cite{andreotti2016open} which simulates noisy abdominal ECG measurements of pregnancies in varying conditions (e.g. fetal movement, uterine contraction) over 34 electrode locations, each recording being 5 minutes at 250Hz. The database contains 10 pregnancies, that differ by the intrinsic characteristics of the mother and the fetus. We use 6 pregnancies for training, 2 for validation, and 2 for testing. Each sample of an electrode provides the noisy ECG mixture and ground-truth for maternal and fetal ECG. All 34 electrodes are given independently to the model, without location information. We train a single model independently of the electrode's location and rely on the same architecture as in our speech experiments, only validating regularization parameters. In this context, the source/speaker stack learns a representation of a mother and its fetus to condition the separation stack. Wavesplit, Dual-Path RNN, and Conv-TasNet are trained on the same data. Table \ref{table:ecg} illustrates the advantage of Wavesplit on this task. A visualization of separated heart rates is shown in Appendix \ref{appendix:fecgsyndb}.

\section{Conclusions}
\label{sec:ccl}

We introduce Wavesplit, a neural network for source separation. From the input mixed signal, our model extracts a representation for each source and estimates the separated signals conditioned on the inferred representations. Contrary to prior work, we learn both tasks jointly in an end-to-end manner, optimizing the reconstruction of the separated signals. For each mixed signal, the model learns to predict local representations which can be aggregated into a consistent representation for each source via clustering. Clustering is well suited for separation as it naturally represents a set of sources without arbitrarily ordering them. Separation with Waveplit relies on a single consistent representation of each source regardless of the input signal length. This is advantageous on long recordings, as shown by our experiments on speech separation. For this competitive application, our model redefines the state-of-the-art on standard benchmarks (WSJ-0-mix and LibriMix), both in clean and noisy conditions. We also report the benefits of Wavesplit on fetal/maternal heart rate separation from electrocardiograms. These results open perspectives in other separation domains, e.g. light curves in astronomy, electrical energy consumption, or spectroscopy.

\vspace{-2mm}
\section*{Acknowledgments}
The authors are grateful to Adam Roberts, Chenjie Gu, Raphael Marinier and Olivier Teboul for their advice on model implementation. They are also grateful to Jonathan Le Roux, John Hershey, Richard F. Lyon, Norman Casagrande and Olivier Pietquin for their help navigating speech separation prior work. The authors also thank Fernando Andreotti and Joachim Behar for the FECGSYNDB dataset.

\vspace{-2mm}
\bibliography{main}
\bibliographystyle{IEEEtran}

\clearpage
\appendices 

{\bf \large Appendix}

\section{Wall Street Journal Mix  (WSJ0-mix) and LibriMix Dataset}
\label{appendix:wsj0mix_stats}

The Wall Street Journal mix dataset (WSJ0-mix) was introduced in~\cite{hershey16:deep_clustering}, while LibriMix was introduced in~\cite{librimix}. 
For LibriMix, we rely on the \textit{train-360} version of the training set.
Table~\ref{table:wsj0mix_stats} reports split statistics for the datasets we used.

\begin{table}[h]
        \centering
        \caption{WSJ0-mix and LibriMix statistics.}
        \vspace{0.2cm}
        \label{table:wsj0mix_stats}
        \begin{tabular}{l|l|r|r|r}
        Dataset   &             & train & valid & test\\\hline
        WSJ0-2mix & \# examples &   20k &    5k &   3k \\
                  & \# speakers &  \multicolumn{2}{c|}{101} & 18 \\
                  & mean length &   5.4 sec & 5.5 sec & 5.7 sec \\\hline
        WSJ0-3mix & \# sequences &  20k & 5k & 3k \\
                  & \# speakers & \multicolumn{2}{c|}{101} & 18 \\
                  & mean length &   4.9 sec & 4.9 sec & 5.2 sec \\\hline
        Libri2mix & \# examples &   51k &    3k &   3k \\
                  & \# speakers &  921 & 40 & 40 \\
                  & mean length &   15.0 sec & 13.2 sec & 13.2 sec \\\hline
        Libri3mix & \# sequences &  34k & 3k & 3k \\
                  & \# speakers & 921 & 40 & 40 \\
                  & mean length &   15.5 sec & 13.2 sec & 13.2 sec \\\hline
        \end{tabular}
\end{table}

\section{Maternal and Fetal Heart Rate Separation}
\label{appendix:fecgsyndb}

\begin{figure*}[h] 
  \begin{center}
  \includegraphics[width=\textwidth]{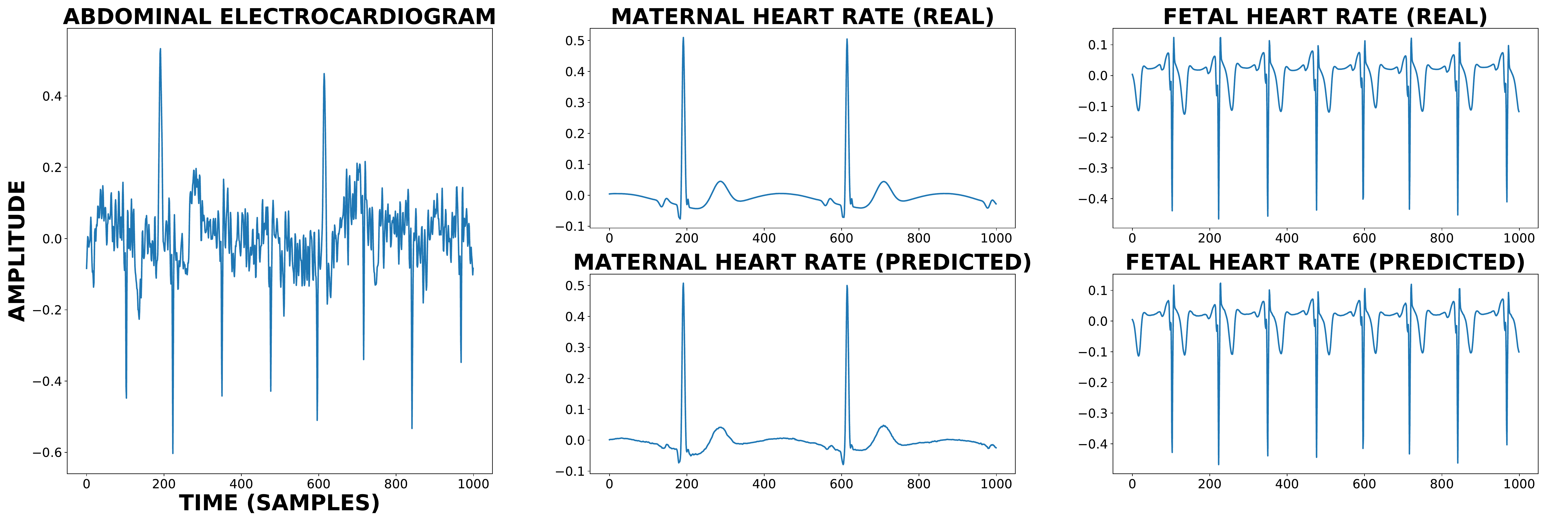}
  \end{center}
  \caption{Example of separation of maternal and foetal heart rate from a simulated abdominal electrode on the FECGSYNDB test set.}
  \label{fig:fecgsyndb}
 \end{figure*}

Figure~\ref{fig:fecgsyndb} provides an example signal of Wavesplit applied to maternal/fetal heart rate separation. The input mixture (left)
shows that the signal is almost indistinguishable from noise outside of peaks. However, our model extracts both the mother (center)
and fetal (right) signal with high accuracy.

\end{document}